\documentclass[a4paper]{article}

\usepackage{icrc2013}
\usepackage[english]{babel}

\title{The H.E.S.S. Galactic Plane Survey - maps, source catalog and
  source population}

\shorttitle{H.E.S.S. survey maps, catalog and source population}

\authors{
S.~Carrigan$^{1}$,
F.~Brun$^{1}$,
R.C.G.~Chaves$^{2,1}$,
C.~Deil$^{1}$,
A.~Donath$^{1}$,
H.~Gast$^{3,1}$,
V.~Marandon$^{1}$,
M.~Renaud$^{4}$
for the H.E.S.S. Collaboration.
}

\afiliations{
$^1$ Max-Planck-Institut f\"ur Kernphysik, P.O. Box 103980, D 69029 \\
$^2$ DSM/Irfu, CEA Saclay, F-91191 Gif-Sur-Yvette Cedex, France \\
$^3$ I. Physikalisches Institut B, RWTH Aachen University \\
$^4$ Laboratoire Univers et Particules de Montpellier, Universit\'e
Montpellier 2, CNRS/IN2P3,  CC 72, Place Eug\`ene Bataillon, F-34095
Montpellier Cedex 5, France \\ 
}

\email{svenja.carrigan@mpi-hd.mpg.de}

\abstract{
The H.E.S.S. Galactic Plane Survey (HGPS), the first comprehensive
survey of the inner Galaxy at TeV energies, has led to the discovery
of an unexpectedly large and diverse population of over 60 sources of
TeV gamma rays within its current range of l = 250 to 65 degrees in
longitude and $|$b$|<$ 3.5 degrees in latitude. The data set of the HGPS
comprises $\sim$2800 hours of high-quality data, taken in the years 2004 to
2013. The sensitivity for the detection of point-like sources is at
the level of 2\% Crab or better 
in the HGPS region. The population of TeV
gamma-ray emitters is dominated by the pulsar wind nebula and
supernova remnant source classes, although nearly a third of the
sources remain unidentified or ambiguous. We are presenting
the latest HGPS significance and sensitivity
maps, as well as a work on the HGPS source catalog, based on a uniform
re-analysis of the full data set collected in the last decade. We will
also give a brief overview of the H.E.S.S. Galactic source population.
}


\begin{document}
\maketitle

\section{Introduction}

The High Energy Stereoscopic System (H.E.S.S.~\cite{bib:Hinton2004}) is
ideally suited for undertaking a deep survey of our Galaxy, due to 
its high sensitivity, comparatively large field-of-view (FoV) of 5$^\circ$, 
and its angular resolution of $\sim$0.1$^\circ$ \cite{bib:Crab06}. Its location in the 
Khomas highlands of Namibia provides it 
a prime view of the inner Galaxy. Here we report on the
status and latest results of the H.E.S.S. Galactic Plane Survey
(HGPS), the deepest and most 
comprehensive survey of the inner Galaxy undertaken in
very-high-energy (VHE, E$>$0.1~TeV) $\gamma$-rays so
far. The latest significance map is shown, alongside a short
introduction to advanced methods for the suppression of cosmic-ray 
background. A software pipeline to detect and model
sources of VHE $\gamma$-rays in the HGPS data set and construct a catalog is introduced. 

The VHE $\gamma$-ray source population is dominated by
objects that are linked to the final stages in stellar evolution,
namely pulsar wind nebulae (PWNe) and supernova remnants (SNRs). For
a large number of sources, however, no plausible counterpart at
other wavelengths has been found yet, or the physical origin of the
detected emission remains unclear. An overview of the
H.E.S.S. Galactic source population, and some of its properties is given.

\section{Maps and source catalog}

The inner Galaxy has now been observed for $\sim$2800~hr
using a variety of observation strategies
\cite{bib:ICRC09,bib:ICRC11} from 2004 to 2012.
The main
goal has always been to discover new VHE $\gamma$-ray sources and enable
studies of populations of Galactic source classes as a consequence.

After calibration and quality selection \cite{bib:Crab06}, a
multivariate analysis 
technique~\cite{bib:Ohm2009} based on shower and image shape parameters
is used to discriminate $\gamma$-ray-like events from
cosmic-ray-induced showers. A minimum image amplitude of 160 photo
electrons is required. 

\begin{figure*}[!t]
  \centering
  \includegraphics[angle=90,width=\textwidth]{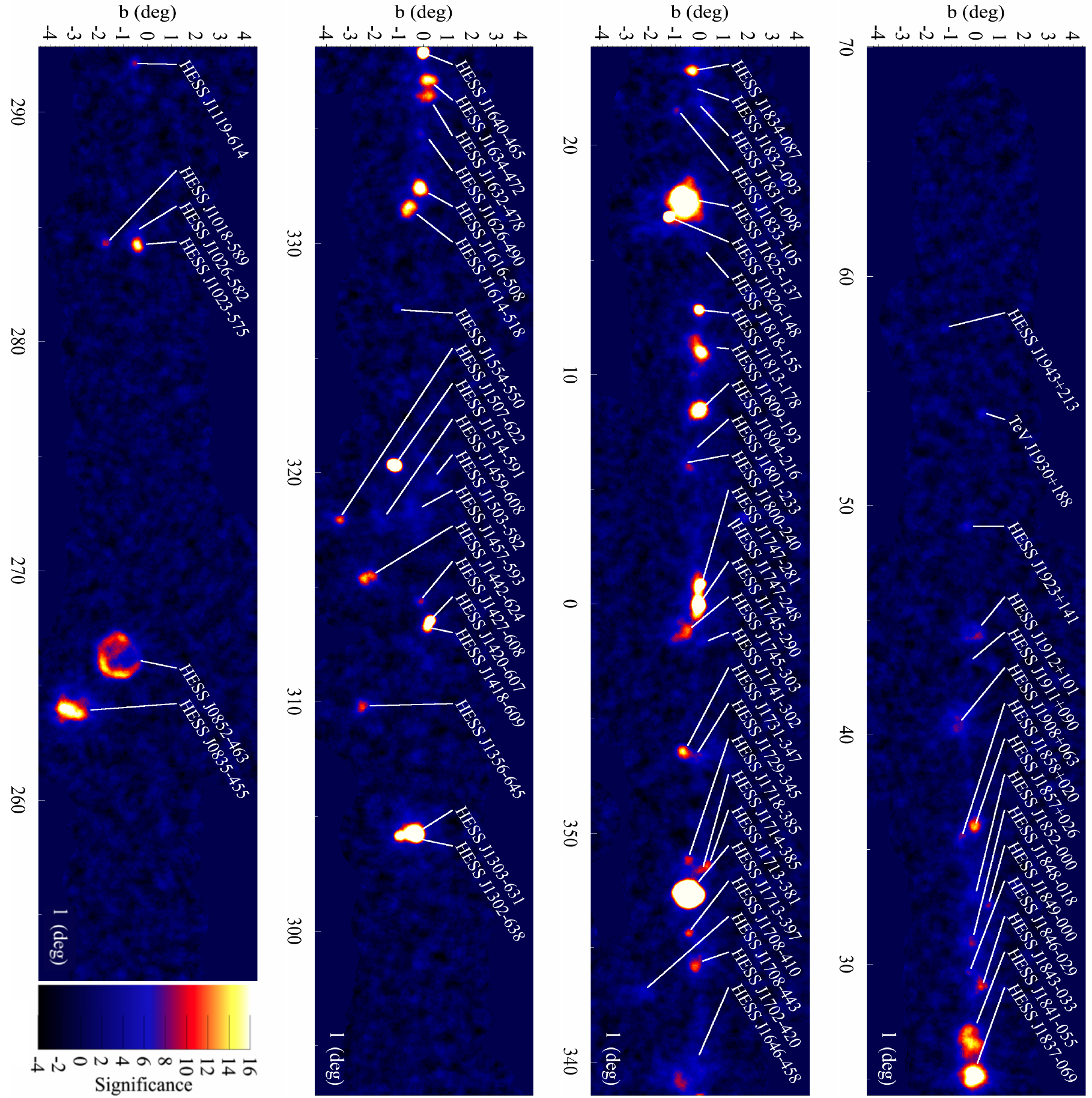}
  \caption{Significance map for the H.E.S.S.~Galactic Plane Survey. The
pre-trials significance for a correlation radius of $0.1$ deg is
shown. The colour transition from blue to red corresponds to
$\sim\!5\sigma$ post-trials significance. The significance has been
calculated for regions on the 
sky where the sensitivity of H.E.S.S.~for point sources ($5\sigma$
pre-trials, and assuming 
the spectral shape of a power law with index 2.3) is better than
10\% Crab. Identifiers for sources that have been described in
publications or announced at conferences are included.}
  \label{sig_fig} 
\end{figure*}

To produce maps, the remaining background
is estimated by the {\it ring background technique}~\cite{bib:Berge2007}, where,
for each map position, the background is estimated from a
ring centered on this position. It is vital to exclude regions with
known  $\gamma$-ray emission from areas in which the background is being
calculated. A procedure has been developed to generate these
\emph{exclusion regions} automatically.
The resulting \emph{exclusion maps} cover areas of the 
sky that are comparatively large on the scale of the size of the FoV.
Therefore, the {\it ring background technique} was modified
to allow the ring radius to adaptively
enlarge when a large fraction of the ring area overlaps with an
excluded region, until an appropriate ring of the same thickness is
reached.  Further details on these two procedures can be found in
\cite{bib:Carrigan2013}.
The significance value for each position is then
calculated~\cite{bib:LiMa1983} by summing the candidate events within a
fixed and predefined radius of e.g. 0.1$^\circ$, suitable 
for point-like sources, and comparing to the estimated background level
at that position. Fig.~\ref{sig_fig} shows the latest
significance map obtained for the survey region using these methods
and the latest data~\footnote{The H.E.S.S. software {\it hap}, version
  {\it hap-13-06}, was used.}.

The current H.E.S.S. sensitivity to VHE $\gamma$-ray sources (here, for
point-like 
sources and assuming a power-law spectrum with index 2.3) is
below 2\% Crab for practically all of the longitude range l = 283$^\circ$  to
59$^\circ$ at latitude b = $-$0.3$^\circ$ (the mean for
known Galactic H.E.S.S. sources) \cite{bib:Carrigan2013}.

Using the HGPS maps as input, we have been developing a
software pipeline to generate a source catalog.
The aim is to have detection criteria as well as morphological and
spectral analyses which are as uniform as possible. 
Such a catalog greatly facilitates population studies of Galactic source classes
as well as measurements of Galactic diffuse emission.
It will be published in the near future alongside the maps and
should also prove useful for the astronomical 
community when compared to data from other wavelengths, from
radio to X-rays to \emph{Fermi}/LAT~\footnote{http://fermi.gsfc.nasa.gov} and
HAWC~\footnote{http://www.hawc-observatory.org}.

To construct the catalog, we use a likelihood fit of the
H.E.S.S. event (count) map, taking the exposure and 
point spread function as well as the estimated background into
account (see \cite{bib:Carrigan2013} for more details on the HGPS catalog
production). 
The measurement of spectra of catalog
sources is done independently from the maps,  
by defining source regions that contain most of the emission from a given
source using the {\it reflected background method}~\cite{bib:Berge2007}.

\section{Galactic source population}
\begin{figure*}[t!]
  \begin{centering}
    \includegraphics[width=0.65\textwidth]{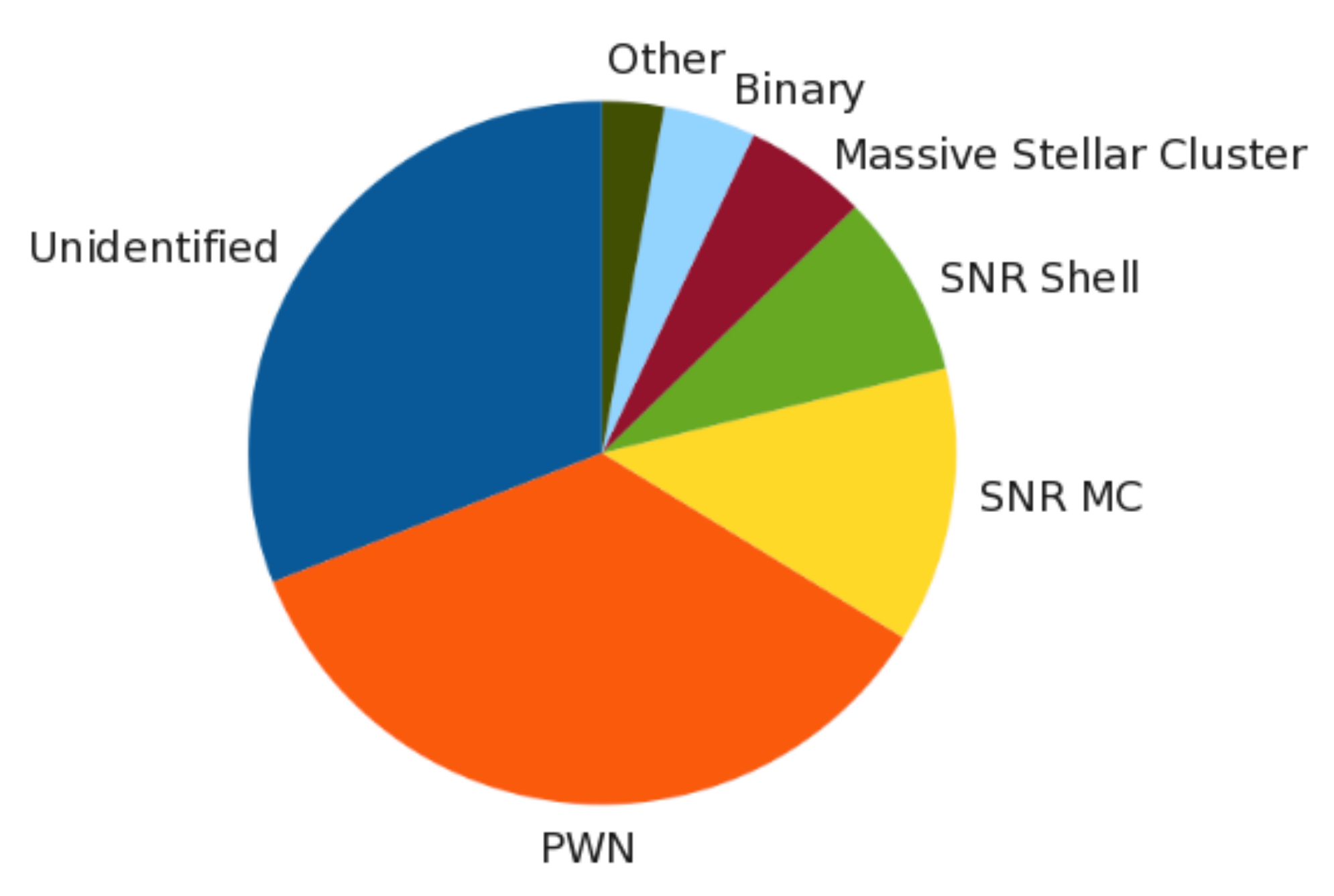}
    \caption
    {Galactic H.E.S.S. sources, status February 2013 (classifications taken from
  TeVCat). Pulsar wind nebulae (PWN, orange) are the largest source class
  with $\sim$35\%, followed by supernova
  remnants, either interacting with a molecular cloud (SNR MC, yellow,
   $\sim$13\%)
  or exhibiting emission from their shell (SNR Shell, light
  green, $\sim$8\%). Besides a few massive stellar clusters (dark red,
  $\sim$6\%) and some  binary systems (light blue, $\sim$4\%) a large
  part of the H.E.S.S. source population remains unidentified (dark
  blue,  $\sim$31\%). 'Other' (dark green) comprises the
  globular cluster Terzan~5 and the high-frequency peaked BL Lac
  object HESS~J1943$+$213.}
    \label{fig:pie_chart}
  \end{centering}
\end{figure*}

Currently, more than 140 sources are listed in the online VHE
$\gamma$-ray catalogue   
TeVCat~\footnote{http://tevcat.uchicago.edu}, of which nearly 70
sources were detected by H.E.S.S. in the Galactic plane. 
The dominant source populations are PWNe and SNRs, but a large
fraction are still unidentified (Fig.~\ref{fig:pie_chart}, data from TeVCat).

\begin{figure*}[!t]
  \begin{centering}
    \includegraphics[width=0.8\textwidth]{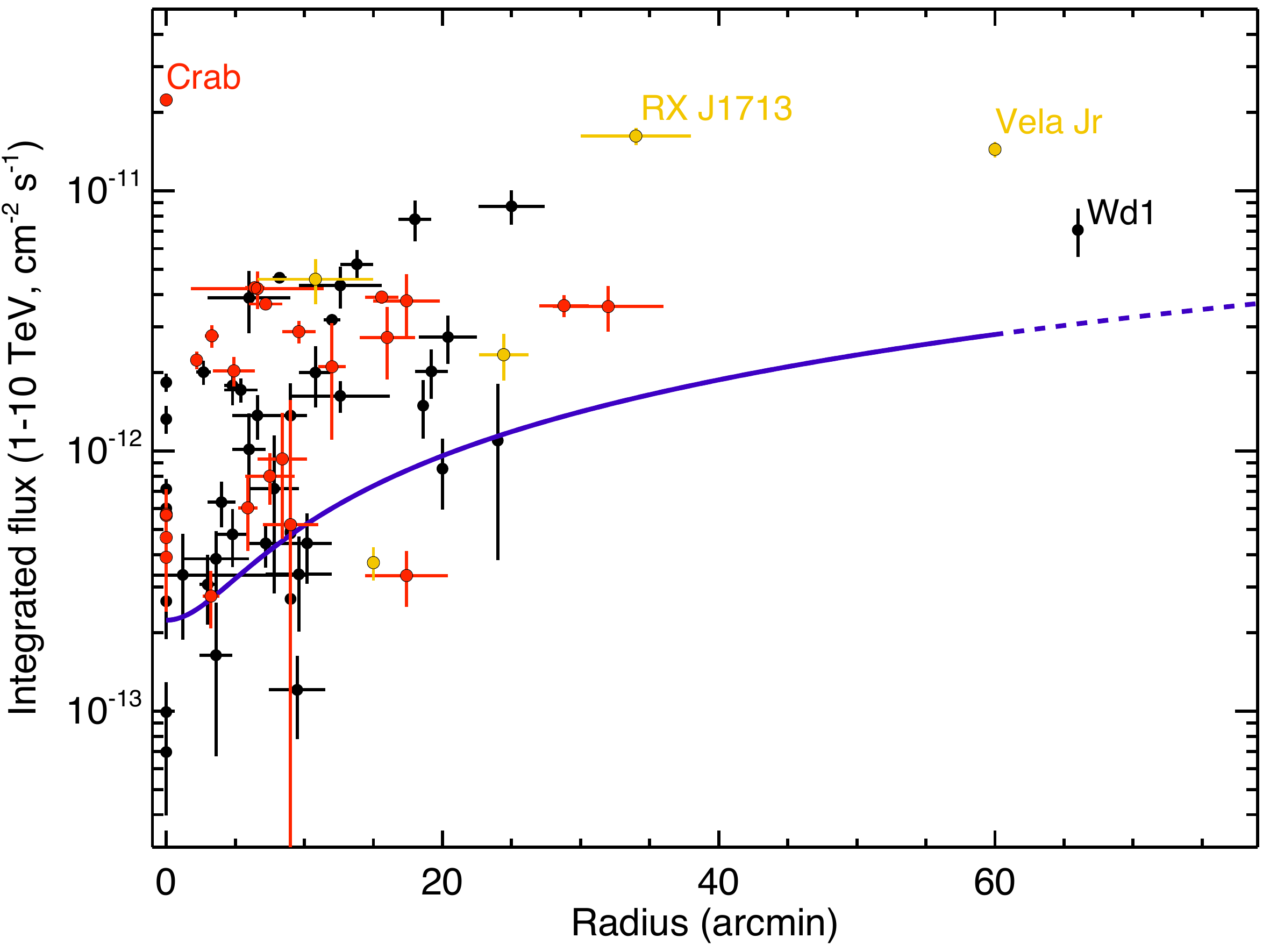}
    \caption
    {H.E.S.S. Galactic sources: Integrated flux in the 1$-$10~TeV
      (cm$^{-2}$ s$^{-1}$) energy band versus source radius (arcmin). The blue
      line depicts an estimate of the radius-dependant sensitivity of
      H.E.S.S.\  SNRs are marked in yellow, PWNe in red, and all other classes 
      (including those unidentified or confused) in black. (Data taken
      from TeVCat.)
    }
    \label{fig:flux_extension}
  \end{centering}
\end{figure*}

The study of source populations needs to be carefully performed,
especially with regards to selection effects which may  
limit our knowledge of the Galaxy.
As an example, the VHE $\gamma$-ray integral flux as a function of the
source radius is shown in   
shown in Fig.~\ref{fig:flux_extension}.  
It reveals that the sample is likely incomplete, with a reduced
sensitivity to large and faint sources. 

\begin{figure*}[!t]
  \centering
  \includegraphics[width=0.8\textwidth]{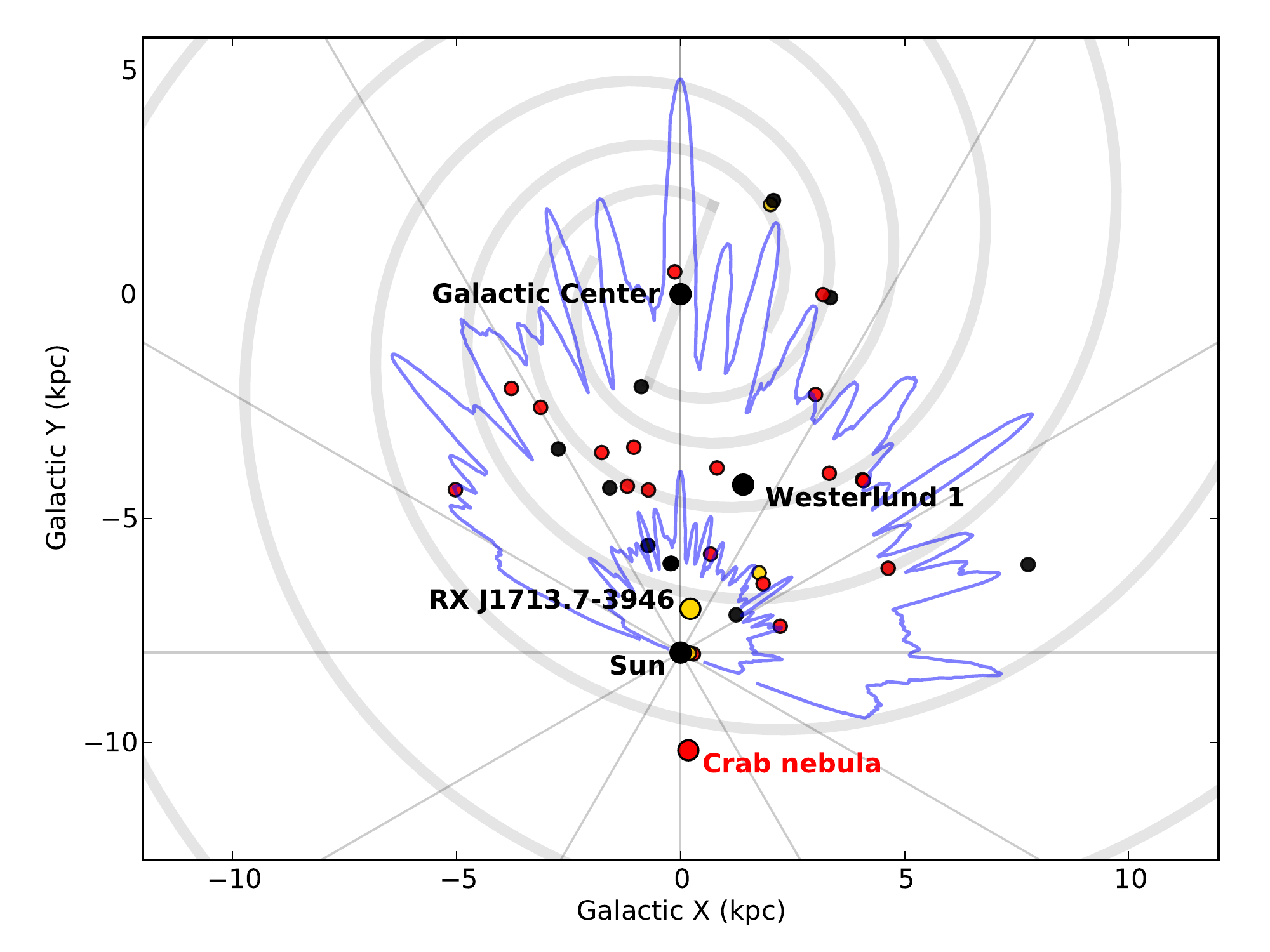}
  \caption
  { Face-on view of our Galaxy, with the spiral arms
    \cite{bib:Vallee2008} drawn as solid grey lines. 
    The H.E.S.S. horizons for 1\% and 10\% of the Crab Nebula
    luminosity above 1 TeV (ph s$^{-1}$), for a 5-$\sigma$ point-like
    source, are depicted by the blue curves.  
    Superimposed are the H.E.S.S. Galactic sources with known
    distances, as listed in TeVCat. 
    Colours as in Fig. 3.
  }
  \label{fig:topview}
\end{figure*}

The question of the completeness of the H.E.S.S. source sample can
also be evaluated in terms of spatial coverage of the Galaxy. 
Figure~\ref{fig:topview} shows a face-on view of our Galaxy, in which
detected sources are plotted at their estimated distances. 
This clearly shows that only a small fraction of the Galaxy is
observed (roughly 50\% of the Galaxy for 10\% Crab luminosity). 
An important caveat to the displayed horizons is that they are
computed based on the sensitivity of the HGPS and do not take into
account the 
screening effect that known sources have in obscuring or confusing
potential emission along the line-of-sight but behind that source. 
Moreover, the horizons are currently calculated only for the
simplified hypothesis of point-like sources. 
As seen in Fig.~\ref{fig:flux_extension}, the sensitivity decreases as
the source size increases, which would yield a closer horizon in the
case of an extended source.  

\section{Outlook}

After nearly a decade of data taking, H.E.S.S. is ending its surveying
program of the Galactic Plane.  The additional of a large 
fifth Cherenkov telesope in the centre of the H.E.S.S. array
marks the beginning of H.E.S.S. Phase II and will concentrate on deeper
observations with much improved intensity and spatial resolution but
smaller FoV.
Here, we have presented an update on the HGPS status and a brief overview of
the H.E.S.S. Galactic source population, including some of its limitations.
In the near future, we aim to present the entire data set
of the HGPS, in the form of high-level maps and a source catalog accessible
to the community.

With this large data set, extensive population studies in the
VHE range become possible for the first time, as will be reported in
this conference in a contribution by S.~Klepser et al., ''A Population
of Teraelectronvolt Pulsar Wind Nebulae in the H.E.S.S. Galactic Plane
Survey'' (ID 591) and in a contribution by J.~Hahn et al., ''Study of
the Very High Energy emission 
from Supernova Remnants with H.E.S.S.'' (ID 1048). Besides population
studies other interesting topics can be explored in this data set,
e.g. the diffuse emission in the Galactic plane, on wich K.~Egberts et al.
will report: ''Diffuse TeV Gamma-Ray Emission in the H.E.S.S. Galactic
Plane Survey'' (ID 744). In-depth studies of the Galactic centre will
be presented by A.~Viana et al., ''Spectral morphology of the inner 50
pc of the Galactic Center region in 
very-high-energy gamma-rays with H.E.S.S'' (ID 901).

\vspace*{0.5cm}
\footnotesize{{\bf Acknowledgment:}{ please see standard
    acknowledgement in H.E.S.S. papers, not reproduced here due to
    lack of space.}} 

%\newpage

\end{document}